\documentclass[twocolumn]{aastex62}

\usepackage{amsmath}

\newcommand{\thi}{$\tau_{\rm HI}$}
\newcommand{\hi}{H\textsc{i}}

\newcommand{\R}{$\mathcal{R}$}
\newcommand{\Rhi}{$\mathcal{R}_{\rm HI}$}
\newcommand{\Rdust}{$\mathcal{R}_{\rm dust}$}
\newcommand{\Rlim}{$1.3$}
\newcommand{\totalLOS}{$151$}
\newcommand{\galfaLOS}{$72$}

\newcommand{\galfaLOShigh}{$51$}

\graphicspath{{./}{figures/}}

\accepted{\today}
\submitjournal{ApJ}

\shortauthors{Murray et al.}

\begin{document}

\title{Optically thick HI does not dominate dark gas in the local ISM}

\correspondingauthor{C.\,E.\,M.}
\email{clairemurray56@gmail.com}

\author[0000-0002-7743-8129]{Claire E. Murray}
\affil{Space Telescope Science Institute, 
3700 San Martin Drive, 
Baltimore, MD, 21218}

\author[0000-0003-4797-7030]{J. E. G. Peek}
\affil{Space Telescope Science Institute, 
3700 San Martin Drive, 
Baltimore, MD, 21218}

\affil{Department of Physics \& Astronomy, 
Johns Hopkins University,
3400 N. Charles Street, 
Baltimore, MD 21218}

\author[0000-0002-9888-0784]{Min-Young Lee}
\affil{Max-Planck-Institut fur Radioastronomie, 
Auf dem H\"ugel 69, 
D-53121 Bonn, Germany}

\author{Sne\v{z}ana Stanimirovi\'c}
\affil{Department of Astronomy, 
University of Wisconsin -- Madison, 
475 N. Charter Street,
Madison, WI 53706}

\begin{abstract}

The local interstellar medium (ISM) is suffused with ``dark" gas, identified by excess infrared and gamma ray emission, yet undetected by standard ISM tracers such as neutral hydrogen (\hi) or carbon monoxide emission. Based on observed dust properties from \emph{Planck}, recent studies have argued that \hi\ mixed with dust is strongly saturated and that dark gas is dominated by optically-thick \hi. We test this hypothesis by reproducing this model using data from \emph{Planck} and new $21\rm\,cm$ emission maps from GALFA-\hi -- the first large-area $21\rm\,cm$ emission survey with comparable angular resolution to \emph{Planck}. We compare the results with those from a large sample of \hi\ column densities based on direct observations of \hi\ optical depth, and find that the inferred column density corrections are significantly lower than those inferred by the \emph{Planck}-based model. Further, we rule out the hypothesis that the pencil-beam \hi\ absorption sight lines preferentially miss opaque ``blobs" with small covering fraction, as these structures require densities and pressures which are incompatible with ISM conditions. Our results support the picture that excess dust emission in the local ISM is not dominated by optically-thick \hi, but is rather a combination of intrinsic changes in dust grain emissivities and H$_2$ missed by CO observations. 
\end{abstract}


\section{Introduction} \label{sec:intro}

Galactic ecosystems rely on the conversion of interstellar medium (ISM) into stars and then back again. Accounting for all components of the ISM is therefore crucial for building a self-consistent model of galaxy evolution. However, there is a significant ISM population that is not traced by standard neutral gas tracers. The prevalence of this ``dark gas" is inferred via gamma ray emission from cosmic ray interactions with neutral gas \citep{2005Sci...307.1292G}, excess far infrared (FIR) emission \citep[e.g.,][]{2011A&A...536A..19P}, dust extinction towards diffuse clouds \citep[e.g.,][]{2012A&A...543A.103P} and ionized carbon emission \citep[e.g.,][]{2014A&A...561A.122L}.

As for what ``dark gas" is made of, one likely culprit is molecular hydrogen (H$_2$), the most abundant interstellar molecule which cannot be observed directly in cold, dense environments. Carbon monoxide (CO) is a popular tracer for H$_2$, as it has strong dipole-allowed rotational transitions which are easily excited at low temperatures. However, CO cannot effectively self-shield and is easily photo-dissociated \citep{1988ApJ...334..771V}, rendering significant portions of H$_2$ clouds untraced by CO emission \citep[e.g.,][]{2010A&A...518A..45L}. 

A second possibility is that dark gas is cold, optically thick \hi\ not detectable via $21\rm\,cm$ emission. This population is difficult to characterize as both $21\rm\,cm$ emission and absorption are required to constrain the excitation temperature (i.e., spin, $T_s$), optical depth (\thi) and total column density of the gas. These measurements require sources of background continuum and careful considerations for radiative transfer effects \citep[e.g.,][]{2003ApJS..145..329H}.

To differentiate between the potential origins of dark gas in light of these observational limitations, indirect tracers are often employed. Assuming that dust and gas are well-mixed, the optical depth of neutral gas (\hi\ and H$_2$) may be traced by observed dust properties. Under the further assumption that the specific dust opacity is constant throughout the ISM, \citet[][hereafter F15]{2014ApJ...796...59F,2015ApJ...798....6F} inferred \thi\ and $T_s$ from \emph{Planck} dust emission for individual clouds and on large scales, and concluded that the ISM is dominated by optically thick \hi\ with $\tau_{\rm HI}\gtrsim 1$. This result implies that local \hi\ mass is $2-2.5$ times greater than what is typically inferred without optical depth correction, with a negligible contribution from CO-faint H$_2$. 

However, direct measurements of \hi\ optical depth typically infer significantly lower optical depth corrections \citep[e.g.,][]{1982AJ.....87..278D,2000ApJ...536..756D,2003ApJ...586.1067H}. For example, \cite{2015ApJ...809...56L} found that optically thick \hi\ measured towards lines of sight (LOS) in and around the Perseus cloud accounts for at most $\sim 20\%$ of dark gas. In agreement, \citet{2015ApJ...811..118R} concluded that atypically high \thi\ is required be consistent with F15 in high-latitude molecular clouds, which is not observed \citep{2017ApJ...834...63R}.

These previous efforts either cover small regions around individual clouds, or have insufficient sample sizes to statistically constrain \hi\ optical depth properties. We will expand these studies to a largely unbiased survey covering a fair fraction of the entire sky. We test the hypothesis that dark gas is optically thick \hi\ by comparing direct measurements of \thi\ with those inferred by reproducing the F15 analysis at high angular resolution. 

\section{Data \& Methods}
\label{s:data}

In this work, we follow the procedures outlined by F15 to predict the contribution of optically thick \hi\ to the total column density from all-sky maps of dust properties from \emph{Planck} and $21\rm\,cm$ emission. All maps are converted to HEALPix\footnote{http://healpix.sf.net/} format \citep{2005ApJ...622..759G} with $N_{\rm side}=1024$ (corresponding to $3.4^\prime$ pixels), and smoothed with a Gaussian beam of FWHM$=4.9^\prime$ to match the all-sky dust maps from \emph{Planck}\footnote{http://irsa.ipac.caltech.edu/Missions/planck.html} (unless they have a lower native resolution). Next, we retrieve a large sample of \thi\ measurements from the literature for comparison.

\subsection{Large-area maps of $21\rm\,cm$ and Dust Emission}

We gather $21\rm\,cm$ emission maps from the Galactic Arecibo L-band Feed Array Survey \citep[GALFA-\hi;][]{2011ApJS..194...20P,2018ApJS..234....2P}. GALFA-\hi\ is the highest angular resolution ($4^\prime$), highest spectral resolution ($0.18\rm\,km\,s^{-1}$) large-area ($13,000\,\rm deg^{2}$) Galactic $21\rm\,cm$ emission survey to date, and the only large area map with comparable angular resolution to \emph{Planck}. 

From the second GALFA-\hi\ data release\footnote{doi:10.7910/DVN/T9CFT8}, we download the all-Arecibo-sky ($0<\alpha_{2000}<360^{\circ}$, $0<\delta_{2000}<35^{\circ}$) \hi\ column density map at Galactic velocities ($-90<v_{\rm LSR}<90\rm\,km\,s^{-1}$)  \citep{2018ApJS..234....2P}. This map has been corrected for stray radiation, or radiation entering the main telescope beam from higher-order sidelobes, via comparison with the meticulously stray-corrected Leiden Argentine Bonn (LAB) survey \citep{2005A&A...440..775K}. We download all GALFA-\hi\ data cubes\footnote{doi:10.7910/DVN/MFM8C7} and construct a map of the peak brightness temperature ($T_{B,\rm peak}$) within the same velocity range. 

To estimate background radio continuum emission ($T_{\rm bg}$), we use the all-sky map of $21\rm\,cm$  emission with $36^\prime$ resolution from the Stockert and Villa-Elisa telescopes \citep{1986A&AS...63..205R, 2001A&A...376..861R}, downloaded from the Centre d'Analyse de Donn\'{e}es Etendues (CADE).\footnote{http://cade.irap.omp.eu/dokuwiki/doku.php?id=stockert}

To trace Galactic dust properties, we use the all-sky map of dust optical depth at $353\rm \, GHz$ ($\tau_{353}$) from the Planck Legacy Archive.\footnote{\url{https://pla.esac.esa.int/pla/}; Version 2.01} This map is the result of modeling dust emission from \emph{Planck} at $353$, $545$ and $857\rm\,GHz$, as well as emission at $100\rm\,\mu m$ from \emph{IRAS} \citep{2014A&A...571A..11P}. 

\subsection{Masking}
\label{s:mask}

As our interest is the contribution of optically thick \hi\ to the total gas column density, we mask regions with significant molecular gas emission. To construct the mask, we use the $^{12}$CO $J=1-0$ integrated intensity map of the Galactic plane at $8.7^\prime$ angular resolution from \citet{2001ApJ...547..792D}. We also include the all-sky \emph{Planck} COMMANDER map of $^{12}$CO $J=1-0$. We mask all pixels with detected CO emission at $>3\sigma$, based on an estimated $\sigma_{\rm CO}=0.3\rm\,K\,km\,s^{-1}$ for the \citet{2001ApJ...547..792D} map and the all-sky error map from \emph{Planck}. 

Following F15, we mask regions with significant ultraviolet radiation which may affect the dust temperature and dust to gas ratio. To construct the mask, we use the all-sky map of $\rm H\alpha$ radiation at $6^\prime$ resolution from \citet{2003ApJS..146..407F} and mask pixels with H$\alpha$ emission $> 5\,\rm Rayleighs$. 

Finally, all Galactic latitudes below $|b|=15^{\circ}$ are masked to avoid multiple components blended in velocity along the LOS. Following F15, we mask regions with significant extragalactic and intermediate velocity emission for which the dust properties may be significantly different, defined by pixels with integrated \hi\ brightness temperature ($W_{\rm HI}$) at $|v_{\rm LSR}|> 100\rm\,km\,s^{-1}$ greater than 10\% of $W_{\rm HI}$ at $|v_{\rm LSR}|<70\rm\,km\,s^{-1}$ and pixels with $W_{\rm HI} > 50\rm\,K\,km\,s^{-1}$ at $35 <|v_{\rm LSR}|< 70\rm\,km\,s^{-1}$ respectively. 

\begin{figure}
\begin{center}
\includegraphics[width=0.5\textwidth]{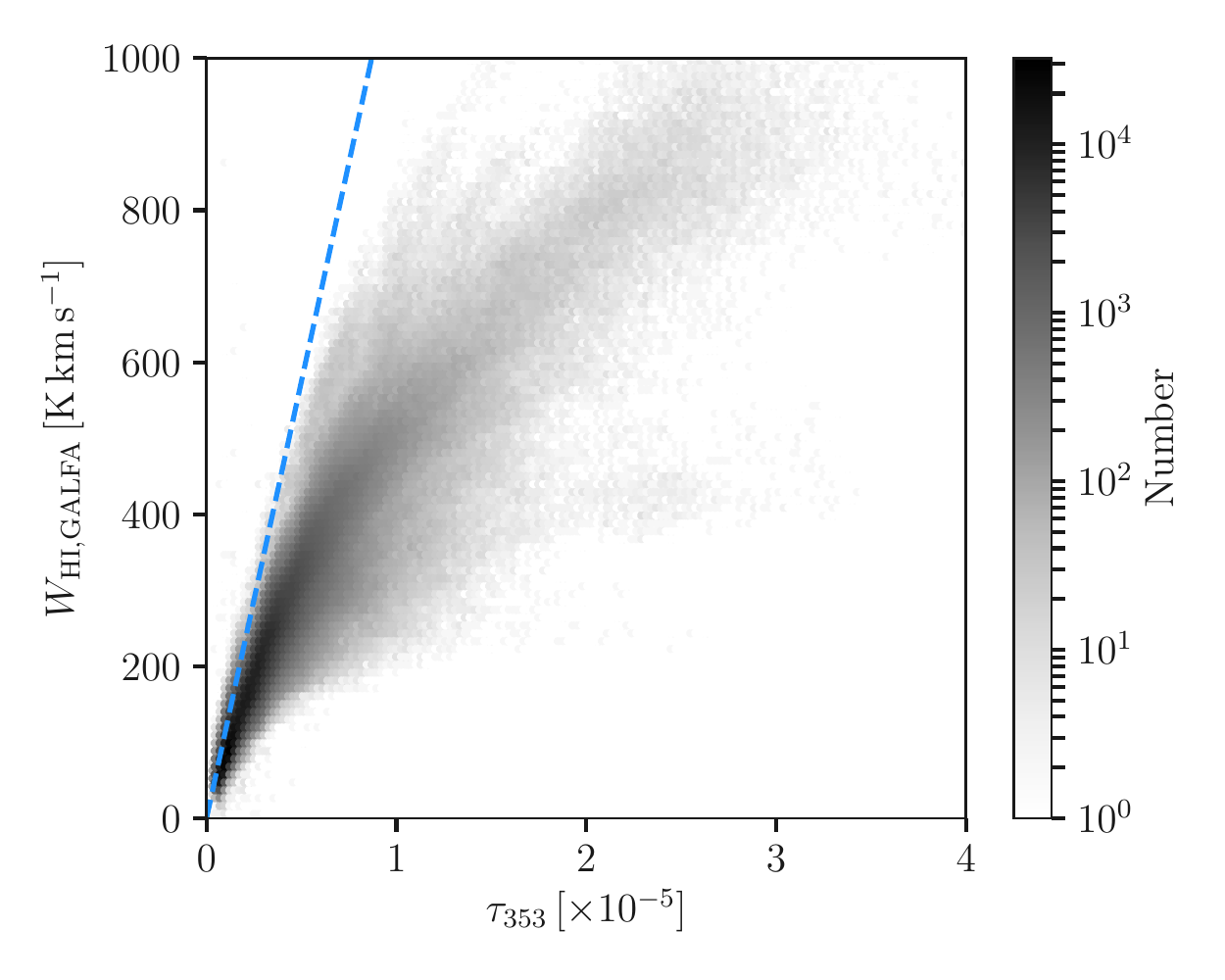}
\caption{Integrated GALFA-\hi\ intensity ($W_{\rm HI}$) vs. $\tau_{353}$ for the unmasked Arecibo sky. The linear relation fitted to data points with highest dust temperatures ($T_d>22.5\rm\,K$) by F15 ($W_{\rm HI} = 1.15\times10^{8}\,\tau_{353}$) is overlaid as a blue dashed line.}
\label{f:hi_tau353}
\end{center}
\end{figure}

\subsection{$21\rm\,cm$ Absorption}

We gather available data from surveys of $21\rm\,cm$ absorption outside of the Galactic plane. The compiled catalog has \totalLOS{} sight lines.

\begin{enumerate}
\item{\emph{21-SPONGE}: We include 57 interferometric \thi\ spectra from a large, high-sensitivity survey for Galactic \hi\ absorption at the Karl G. Jansky Very Large Array: 21-SPONGE \citep[Murray et al.\,2018, submitted]{2015ApJ...804...89M}. 21-SPONGE achieved excellent RMS noise in \thi\ (median $\sigma_{\tau \rm HI} = 0.001$ per $0.42\rm\,km\,s^{-1}$ channels), and includes matching expected emission profiles along the same LOS constructed by interpolating $21\rm\,cm$ emission from the Arecibo Observatory across each target position following the methods of \citet{2003ApJS..145..329H}.}

\item{\emph{Roy et al.\,2013}: We include interferometric \thi\ data presented in \citet{2013MNRAS.436.2352R} from the GMRT, WSRT and ATCA, with matching $21\rm\,cm$ emission spectra from LAB. In the absence of publicly-available spectra, we extract data from their Table 1, including \hi\ column densities under the optical depth-corrected and optically-thin approximations. We select the 22 LOS which are unique relative to 21-SPONGE.}

\item{\emph{The Millennium Survey}: We include single-dish \thi\ spectra from the publicly available Millennium Arecibo $21\rm\,cm$ Absorption-Line Survey \cite[][hereafter HT03]{2003ApJS..145..329H, 2003ApJ...586.1067H}. The median sensitivity  is $\sigma_{\tau \rm HI} = 0.01$ per $0.18\rm\,km\,s^{-1}$ channels. We select the 47 LOS unique in comparison with 21-SPONGE and \citet{2013MNRAS.436.2352R}, because HT03 is generally less sensitive and interferometric absorption measurements filter out contamination from $21\rm\,cm$ emission within the beam \citep[however, we note excellent correspondence between SPONGE and HT03 where they overlap;][]{2015ApJ...804...89M}.}

\item{\emph{Stanimirovi\'c et al. 2014}: We include single-dish \thi\ data from the Arecibo study of the Perseus cloud by \citet{2014ApJ...793..132S}, observed in the same manner as HT03. We gather data from Figure 5 of \citet{2015ApJ...809...56L}, selecting the 25 LOS with unique relative to 21-SPONGE, \citet{2013MNRAS.436.2352R} and HT03.}
\end{enumerate}

\begin{figure*}
\begin{center}
\includegraphics[width=1.0\textwidth]{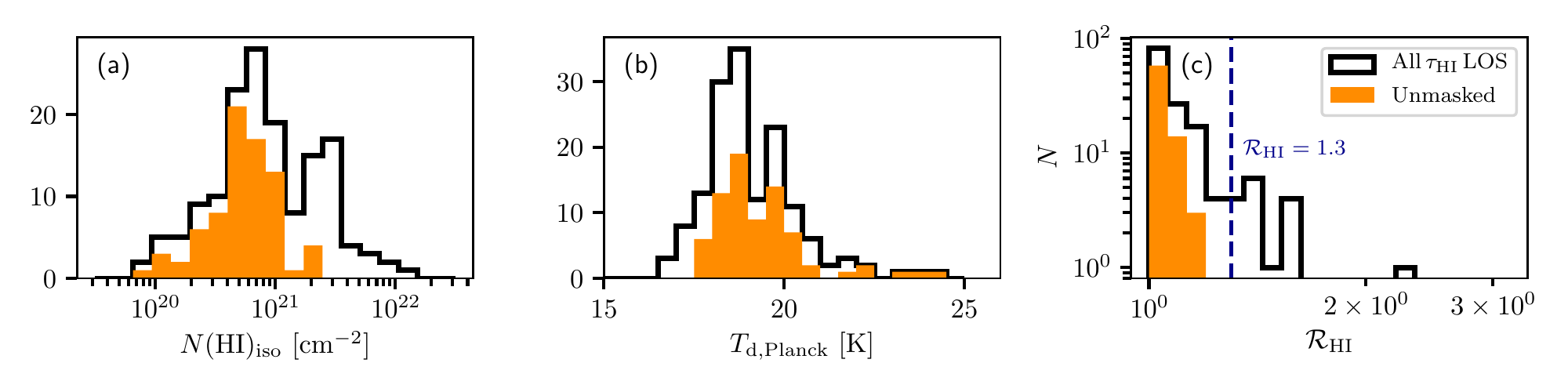}
\caption{Histograms of ISM properties in the direction of \totalLOS{} \thi\ LOS (solid black) and the subset of \galfaLOS{} LOS (filled orange) within the unmasked GALFA-\hi\ FOV. (a) \hi\ column density under the isothermal approximation: $N({\rm HI})_{\rm iso}$ (Equation~\ref{e:nhi_iso}); (b) \emph{Planck} dust temperature: $T_{d,\rm dust}$; (c) \Rhi\ (Equation~\ref{e:r_hi}). Vertical line in panel (c) denotes \Rhi$=1.3$, a limit further discussed in Section~\ref{s:results} and~\ref{s:discussion}}. 
\end{center}
\label{f:hi_props}
\end{figure*}

\section{Analysis}
\label{s:analysis}

The total \hi\ column density, $N({\rm HI})$, is given by,

\begin{equation}
N({\rm HI}) = C_0 \,\int \tau_{\rm HI}\,\, T_s\,\, dv , 
\label{e:nhi}
\end{equation}

\noindent where $C_0 = 1.823 \times10^{18}\rm\,cm^{-2}/(K\,km\,s^{-1})$ \citep[e.g.,][]{2011piim.book.....D}. Under the assumption that all gas along the LOS is a single temperature, $N({\rm HI})$ is given by,

\begin{equation}
N({\rm HI})_{\rm iso} = C_0 \int \frac{\tau_{\rm HI} \,\,T_B}{(1-e^{-\tau_{\rm HI}})}\,\, dv .
\label{e:nhi_iso}
\end{equation}

\noindent where the subscript ``iso" denotes the isothermal approximation \citep[e.g.,][]{1982AJ.....87..278D}. In the optically-thin limit ($\tau_{\rm HI}\ll 1$), Equation~\ref{e:nhi_iso} reduces to,

\begin{equation}
N({\rm HI})_{\rm thin} = C_0\,\int T_B \, dv = C_0\,\, W_{\rm HI}, 
\label{e:nhi_thin}
\end{equation}

Equation~\ref{e:nhi_thin} is used to approximate $N({\rm HI})$ in the absence of optical depth information. The ratio of the total and optically-thin column densities, $\mathcal{R} = N({\rm HI})/N({\rm HI})_{\rm thin}$, is a measure of the contribution of optically thick \hi\ to the \hi\ mass budget.

\subsection{Inferring \R\ via $21\rm\,cm$ absorption}

Observations of both $21\rm\,cm$ emission and absorption are required to constrain \thi\ and $T_s$ for measuring $N({\rm HI})$. Furthermore, the complex velocity structure of Galactic $21\rm\,cm$ spectra implies that clouds along the LOS may have different properties, rendering the isothermal approximation in Equation~\ref{e:nhi_iso} invalid. Significant efforts have been made to extract \thi, $T_s$ and $N({\rm HI})$ for individual clouds along the LOS \citep[e.g.][]{2003ApJ...585..801D,2003ApJ...586.1067H,2004JApA...25..185M,2013MNRAS.436.2352R,2015ApJ...804...89M,2017ApJ...837...55M}. 

\begin{figure*}[t!]
\begin{center}
\includegraphics[width=1.0\textwidth]{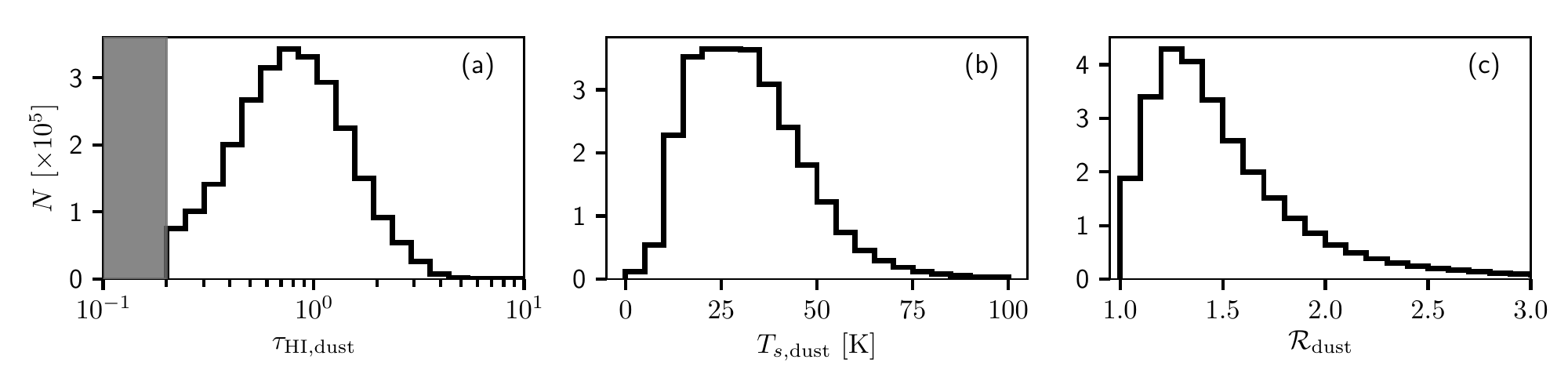}
\caption{Histograms of \hi\ properties inferred from all-sky dust and $21\rm\,cm$ emission, following the analysis of \cite{2014ApJ...796...59F,2015ApJ...798....6F}. (a) $\tau_{\rm HI,dust}$; (b) $T_{s,\rm dust}$; (c) \Rdust (Equation~\ref{e:r_dust}).}
\end{center}
\label{f:tau_ts}
\end{figure*}

Fortunately, \R\ inferred from multi-component analysis has been shown to be statistically indistinguishable from that inferred for the isothermal approximation (Equation~\ref{e:nhi_iso}) \citep[e.g., ][Murray et al.\,2018, submitted]{2015ApJ...809...56L,2018arXiv180511787N}. \citet{2015ApJ...809...56L} concluded that the estimates are equivalent as a result of low observed \thi\ by existing \hi\ absorption studies. Via Monte Carlo simulations of the multiphase ISM, \citet{2013MNRAS.432.3074C} found that $N({\rm HI})_{\rm iso}$ traces the true \hi\ column density for $N({\rm HI})<5\times10^{23}\rm\,cm^{-2}$ regardless of the temperature distribution along the LOS. This indicates that $N({\rm HI})_{\rm iso}$ traces \R\ accurately and with smaller uncertainty than from LOS decomposition. Therefore, for datasets where \thi\ is available, we compute $N({\rm HI})_{\rm iso}$  and $N({\rm HI})_{\rm thin}$ and gather published values where spectra are not available \citep[e.g.,][]{2013MNRAS.436.2352R}. We define \Rhi\ for each LOS via,

\begin{equation}
\mathcal{R}_{\rm HI} = N({\rm HI})_{\rm iso} / N({\rm HI})_{\rm thin}.
\label{e:r_hi}
\end{equation}

\noindent The uncertainty in \Rhi\ is propagated from uncertainties in $N({\rm HI})_{\rm iso}$ and $N({\rm HI})_{\rm thin}$ estimated from the RMS noise in off-line channels of $T_B$ and \thi.

In Figure ~\ref{f:hi_props} we display histograms of $N({\rm HI})_{\rm iso}$, $T_d$ and \Rhi\ for all \totalLOS{} \thi\ LOS (black, unfilled) and for the \galfaLOS{} LOS in the unmasked Arecibo sky (orange, filled). The LOS probe a wide range of column densities and dust temperatures. We note that for all LOS, $N({\rm HI})_{\rm iso} \ll 5\times10^{23}$, validating the use of $N({\rm HI})_{\rm iso}$ as a reasonable approximation of the total \hi\ column density \citep[e.g.,][]{2013MNRAS.432.3074C}.

\subsection{Inferring \R\ via dust properties}

Unfortunately, \Rhi\ measurements are limited by source availability and probe only the gas subtended by the angular size of the source (typically $\ll 1^\prime$). To quantify the contribution of \thi\ to $N({\rm HI})$ over large areas, indirect measures of the optical depth properties of gas are required.

In Figure~\ref{f:hi_tau353} we plot integrated \hi\ intensity from GALFA-HI ($W_{\rm HI,GALFA}$) versus $\tau_{\rm 353}$ for the unmasked sky. \citet{2014ApJ...796...59F,2015ApJ...798....6F} observed that the correlation between $W_{\rm HI}$ and $\tau_{353}$ grows stronger for increasing dust temperature ($T_d$), and fit a linear model to points with $T_d>22.5\rm\,K$. We include the same fit in Figure~\ref{f:hi_tau353} ($W_{\rm HI} = 1.15\times10^8 \cdot \tau_{353}$; F15), which is consistent with our results. F15 hypothesized that highest $T_d$ correspond to optically-thin \hi, supported by evidence that $T_d$ increases with decreasing gas column density \citep{2014A&A...571A..11P}. F15 further asserted that, assuming that dust properties are uniform throughout the sky, this relation holds even if the \hi\ is not optically thin, so that the total \hi\ column density is given by,

\begin{equation}
N({\rm HI})_{\rm HI, dust} = C_0 \cdot k \cdot \tau_{353} = 2.1\times10^{26} \cdot \tau_{353}, 
\end{equation}

\noindent where $k=1.15\times10^{8} \,\rm cm^{-2}$ (F15). 

Using this relation, we repeat the F15 procedure of solving coupled equations (their Equations 4 and 6) to estimate $T_s$ and $\tau_{\rm HI}$ from $\tau_{\rm 353}$, $N({\rm HI})_{\rm thin}$, $T_{\rm bg}$ and $T_{B,\rm peak}$ for each pixel via least squares fit. As noted by F15, this method is valid for $\tau_{\rm HI} \gtrsim 0.2$. We denote the resulting \thi\ and $T_s$ estimates, which represent average conditions, as $\tau_{\rm HI,dust}$ and $T_{s,\rm dust}$ respectively. Finally, we compute the correction to the optically-thin column density, \Rdust, following F15 (their Equation 5) via,

\begin{figure*}[t!]
\begin{center}
\includegraphics[width=\textwidth]{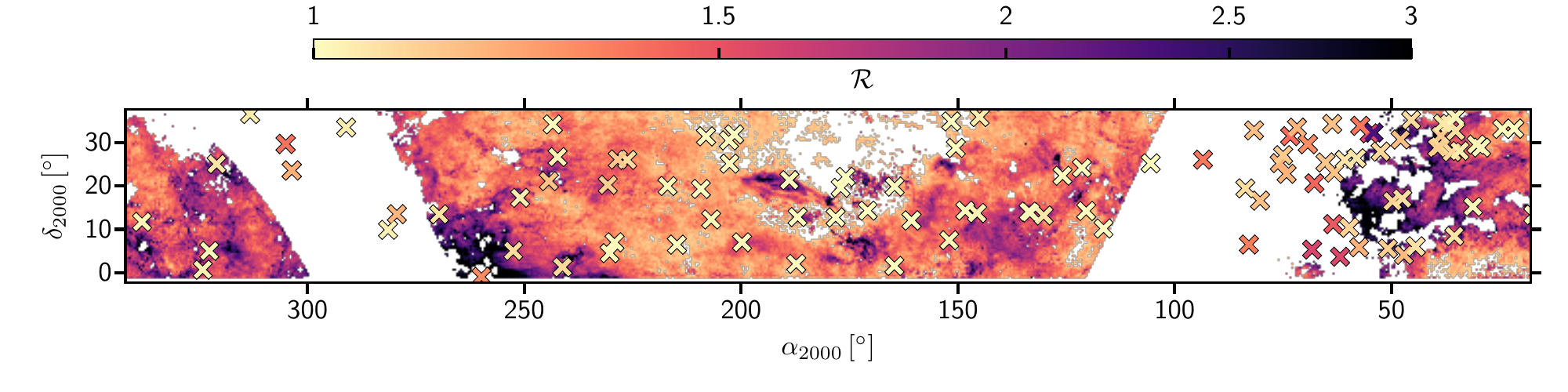}
\caption{Map of \Rdust\ (Equation~\ref{e:r_dust}) for the full GALFA-HI sky, masked according to Section~\ref{s:mask}. Pixels with undefined \Rdust\ (i.e., $\tau_{\rm dust} \leq 0.2$) are also masked. The positions of the 121/\totalLOS{} available \thi\ LOS within the GALFA-HI FOV are overlaid as crosses with colors corresponding to \Rhi\ (Equation~\ref{e:r_hi}).}
\end{center}
\label{f:r_dust_map}
\end{figure*}

\begin{equation}
\mathcal{R}_{\rm dust} = \frac{\tau_{\rm HI,dust}}{(1 - e^{-\tau_{\rm HI,dust}})}.
\label{e:r_dust}
\end{equation}

\noindent Uncertainty in \Rdust\ is estimated by propagating observational uncertainties in $\tau_{353}$ and $N({\rm HI})_{\rm thin}$ from \emph{Planck} and GALFA-HI.

\section{Results}
\label{s:results}

In Figure~\ref{f:tau_ts} we display histograms of $\tau_{\rm HI,dust}$, $T_{s,\rm dust}$ and \Rdust. The results generally agree with F15 (c.f., their Figures 7 and 13). By eye, our distributions appear skewed towards smaller $\tau_{\rm HI,dust}$, larger $T_{s,\rm dust}$ and smaller \Rdust. With the inclusion of the all-sky \emph{Planck} CO map, we mask larger areas of sky with significant CO emission, and therefore the F15 results are likely contaminated at some level by CO-bright gas.

We display a map of \R\ in Figure~\ref{f:r_dust_map}. The map is colored by \Rdust, with \Rhi\ for the 121 LOS within the GALFA-\hi\ field of view (FOV) overlaid. Large regions feature \Rdust$ > 1.5$, corresponding to $\tau_{\rm HI,dust} >1$. 
However, even at low latitudes, within the masked regions of the \Rdust\ map (Figure~\ref{f:r_dust_map}), the correction inferred from \hi\ is $\mathcal{R}_{\rm HI}\lesssim 1.3$. \footnote{We note that saturated \thi\ was measured towards one \citet{2014ApJ...793..132S} source ($\rm 4C+32.14$) and two 21-SPONGE sources ($\rm PKS1944+251$, $\rm J2021+3731$, excluded from their catalog). For these LOS, \Rhi$\gtrsim 2$. However, these LOS are masked due to significant CO emission or low latitude (Section~\ref{s:mask}).}

\begin{figure}
\begin{center}
\includegraphics[width=0.5\textwidth]{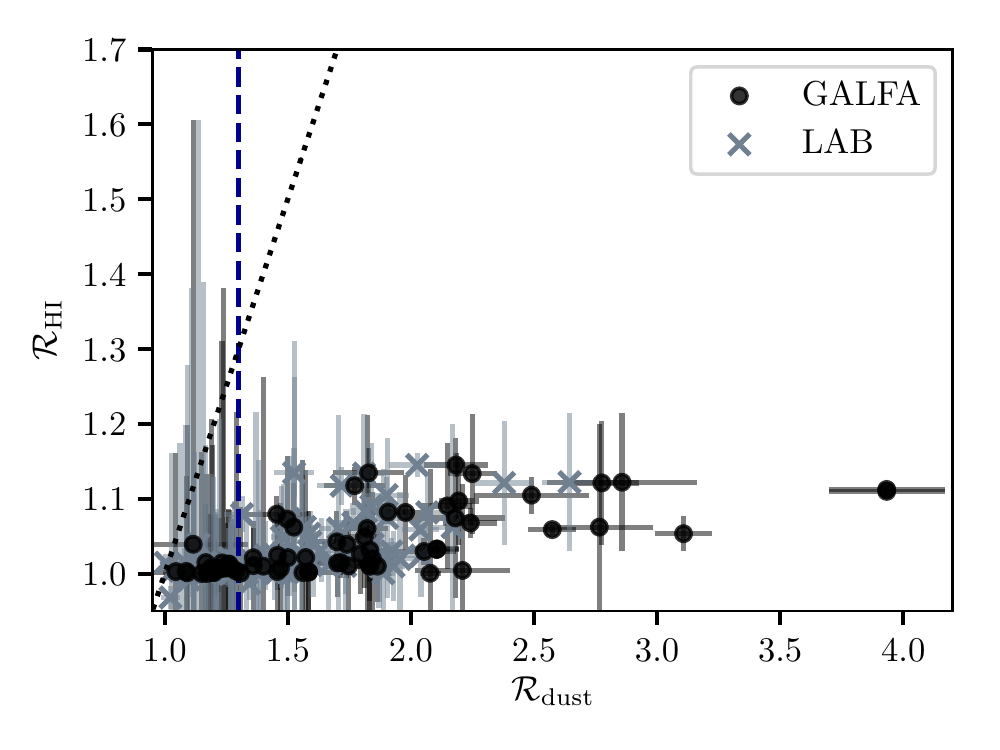}
\caption{Comparing \R\ from direct measurement of \thi\ (\Rhi; Equation~\ref{e:r_hi}) for the \galfaLOS{} \thi\ LOS within the unmasked GALFA-\hi\ sky, and inferred from dust emission (\Rdust; Equation~\ref{e:r_dust}). \Rdust$=$\Rlim, above which we observe no \Rhi, is plotted as a blue dashed line. Gray crosses illustrate the results of a parallel analysis of the same data at LAB survey resolution.}
\end{center}
\label{f:r_factor}
\end{figure}

In Figure~\ref{f:r_factor} we compare \Rdust\ and \Rhi\ for the \galfaLOS{} LOS in the unmasked sky. Although \Rhi\ and \Rdust\ are consistent for some LOS below \Rdust$=$\Rlim, above this value we have \Rdust$>$\Rhi. To be consistent with F15, we repeat the analysis for the same datasets at LAB resolution ($N_{\rm side}=128$, corresponding to $27.5^\prime$ pixels) and with $21\rm\,cm$ emission from LAB, and include the results in Figure~\ref{f:r_factor} (gray crosses). We include an all-sky map of \Rdust\ from our analysis of LAB data in Appendix Figure~\ref{f:r_dust_map_lab}. We find consistent \Rdust\ from LAB as from GALFA-\hi, indicating that our results not biased by the Arecibo FOV. Considering the difference in angular resolution between GALFA-\hi\ ($4^\prime$) and LAB ($36^\prime$), the coherence between results in Figure~\ref{f:r_factor} suggests that \hi\ is largely diffuse at high latitudes down to on $4^\prime$ scales and the lower angular resolution observations are not missing significant unresolved \hi\ contrast. 

Furthermore, although accurately measuring the column density of ``dark" gas requires careful modeling of both gamma ray and FIR emission \citep[e.g.,][]{2015A&A...582A..31P}, and an all-sky map is not yet available to our knowledge, we find that where dark gas maps are available \citep{2017A&A...601A..78R}, specifically in the anticentre and Chameleon molecular cloud regions, regions of \Rdust$\sim 2-3$ correspond to significant dark gas column densities ($N({\rm H})_{\rm dark} \gtrsim 10^{20}-10^{21}$). Figure~\ref{f:r_dust_map_lab}, in Galactic coordinates, may be compared with existing dark gas maps \citep[e.g.,][]{2005Sci...307.1292G}.

\section{Discussion}
\label{s:discussion}

We observe that LOS with significant \hi\ saturation inferred from dust (i.e., \Rdust$\sim 2$) feature \hi\ optical depth corrections ranging from \Rhi$=1.0$ to $\sim 1.3$. If these LOS are indicative of the average \Rhi, then we have ruled out the F15 hypothesis that optically-thick \hi\ comprises the majority of dark gas in the local ISM probed by our high-latitude LOS.

But what if our \thi\ LOS miss optically-thick gas with small covering fraction? For example, \citet{2017arXiv170107129F} asserted that the solid angle coverage of \hi\ emission/absorption observations is too small to sample highly-filamentary, cold \hi. For \Rdust$<1.3$, \Rhi\ and \Rdust\ are often consistent within uncertainties and are therefore not useful discriminants. However, we find that all \galfaLOShigh{} LOS with \Rdust$>$\Rlim\ have \Rhi$<$\Rlim. 

In the following, we compute what the properties of missing optically thick \hi\ structures --- ``blobs" for want of a more descriptive term --- must be to account for the discrepancy between \Rdust\ and \Rhi. From Poisson statistics, the probability of observing \Rhi$>$\Rlim\ zero times in \galfaLOShigh{} LOS is $<10\%$ with $99\%$ certainty. We conservatively assume that this value ($10\%$) is the maximum blob covering fraction. A \emph{Planck} $353\rm\, GHz$ pixel is 4.9$^\prime$ across, and therefore a blob covering 10\% of a pixel area is at most $1.55^\prime$ across. Each blob must account for the missing $N({\rm HI})$ in 10\% of the area, and therefore the column density through a blob must be ten times higher than for 100\% covering factor. For the \galfaLOShigh{} LOS with \Rdust$>$\Rlim, these blobs have necessary column densities from $8\times 10^{20}$ cm$^{-2}$ to $8\times 10^{22}$ cm$^{-2}$ with a mean value of $8\times 10^{21}$ cm$^{-2}$. As distances are difficult to determine, we use a conservative estimate of the \hi\ scale height of $200\rm\, pc$ \citep{1990ARA&A..28..215D} to compute a typical blob distance of $200/\sin{|42^\circ|} = 300\rm\,pc$, where $42^\circ$ is the median blob latitude. Assuming the blobs have no preferred orientation and that they are no deeper than they are across, their diameters are $0.13 \rm\,pc$ with an average proton density of $19,000 \rm\, cm^{-3}$ and volume filling fraction of $<4\times10^{-4}$. To estimate the minimum pressure of these blobs, we assume the coldest known temperature for Galactic \hi-only clouds:  $17\rm\,K$ \citep[the Local Leo Cold Cloud;]{2003ApJ...586.1067H,2011ApJ...735..129P}. Assuming an ideal gas in local thermodynamic equilibrium, this temperature yields a minimum blob pressure of $P/k_B = 3.2 \times 10^{5} \rm \,K\, cm^{-3}$. 

\begin{figure}
\begin{center}
\includegraphics[width=0.5\textwidth]{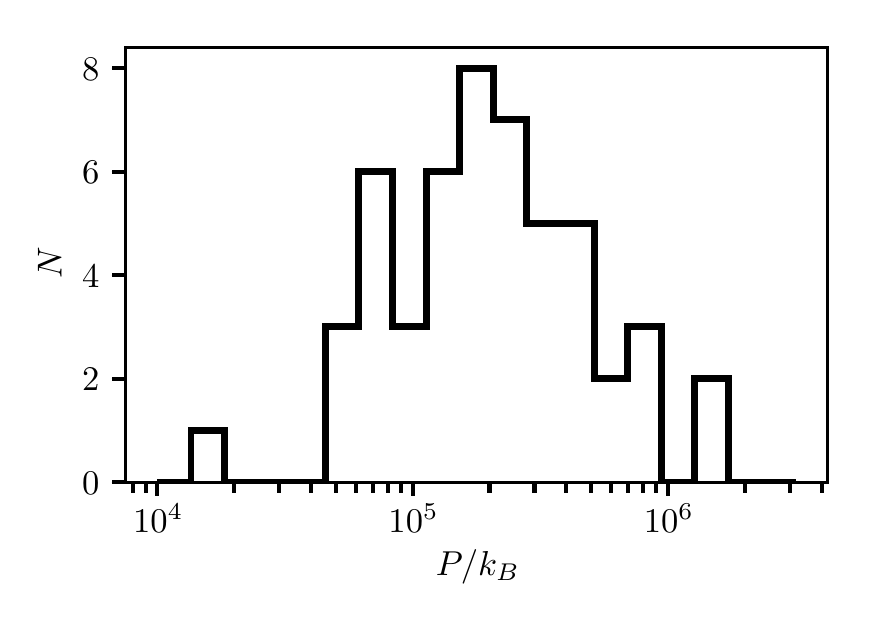}
\caption{Histogram of minimum inferred pressures for opaque \hi\ ``blobs'' required to explain the discrepancy between \Rhi\ and \Rdust\ for the 52 LOS with \Rdust$ > 1.3$ and \Rhi$<1.3$. }
\end{center}
\label{f:blob_pressure}
\end{figure}

In Figure~\ref{f:blob_pressure} we display a histogram of the minimum inferred pressures for all 52 blobs under the same assumptions as above. The observed mass fraction of \hi\ at these high pressures (i.e., $\gtrsim 10^5$) is exceedingly small ($\sim 0.05\%$) and typical pressures are orders of magnitude lower, $\sim 3800$ K cm$^{-3}$ \citep{2011ApJ...734...65J}. So, the blobs should expand to reach pressure equilibrium with their surroundings within a sound-crossing time, $\sim 20,000$ years, without some containment mechanism.

Although ``tiny scale atomic structures" (TSAS) have been inferred from opacity variations on tens of $\rm AU$ scales with $\Delta \tau_{\rm HI}\lesssim0.5$ \citep[e.g.,][]{2005AJ....130..698B, 2009AJ....137.4526L}, the required blob column densities to explain \Rdust$>1.3$ are orders of magnitude higher than observed for TSAS, and these structures must be short-lived with small overall mass fraction \citep[e.g., $<10\%$;][]{1990ARA&A..28..215D}. Furthermore, although geometrical arguments --- for example, the end-on alignment of curved filaments or sheets --- may explain anomalous inferred properties of TSAS \citep[e.g.,][]{1997ApJ...481..193H}, the same arguments break down under the requirement that blob column densities are $100\times$ larger than for typical TSAS and $\sim10\times$ larger than for cold neutral \hi\ (typically $\sim 10^{20}\rm\,cm^{-2}$). 

So, how do we account for the discrepancy between \Rhi\ and \Rdust? It is highly likely that significant variations in dust grain properties render the F15 assumption of uniform dust opacity invalid. On one hand, the dust opacity assumed by F15 ($\tau_{353}/N({\rm HI}  = 4.8\times10^{-27}$; e.g., Equation 5) is $\sim 30\%$ smaller than found by \emph{Planck} for high-latitude cirrus \citep[$7.1\pm0.6\times10^{-27}$;][]{2014A&A...566A..55P}, corresponding to a $\sim 30\%$ increase in their inferred \Rdust. Although F15 found that allowing $\tau_{353}$ to vary as $N({\rm HI})^{1.28}$ based on \emph{Herschel} observations of Orion A \citep{2013ApJ...763...55R} did not affect their conclusions, it is unclear that this relation should hold for the diffuse ISM. From a careful selection of optical-depth corrected \hi\ sight lines with no molecular gas (defined by the non-detection of CO or OH emission at high sensitivity), \citet{2018arXiv180511787N} found that the specific dust opacity varies by up to $\sim 40\%$, likely as a result of dust grain evolution. In support, from detailed studies of the total gas column densities ($N({\rm H})$) traced by gamma ray and dust emission, \citet{2015A&A...582A..31P} and \citet{2017A&A...601A..78R} observed a systematic increase in dust opacity as $N({\rm H})$ increases between the diffuse atomic, ``dark'' and CO-bright phases in the Chameleon and anticentre clouds. This has been predicted theoretically \citep[e.g.,][]{2015A&A...577A.110Y} and inferred by \emph{Planck} dust models \citep[e.g.,][]{2016A&A...586A.132P}, which demonstrated that variations between FIR and optical dust grain properties are incompatible with standard models and require both varying radiation fields and optical grain properties \citep{2015A&A...579A..15K, 2015A&A...580A.136F}. 

In addition, H$_2$ undetected by CO should be prevalent. First of all, the F15 mask threshold of $I({\rm CO}) \lesssim 1\rm\,K\,km\,s^{-1}$ does not guarantee that gas is H$_2$-free. In the diffuse ISM, CO has been detected in emission \citep[$I({\rm CO}) < 1\rm\,K\,km\,s^{-1}$;][]{2012A&A...541A..58L} and via UV and millimeter absorption \citep[e.g.,][]{2007ApJS..168...58S,2008ApJ...687.1075S}. Furthermore, in massive high-latitude clouds, neither optically-thick \hi\  nor variations in dust grain emissivity are sufficient to explain the observed wide range of dust emission per unit gas column density, indicating that H$_2$ must contribute \citep{2017ApJ...851..119R,2017ApJ...834...63R}. In addition \citet{2014ApJ...783...17L} observed significant flattening of $N({\rm HI})/E(B-V)$ at $E(B-V)\gtrsim0.1\rm\,mag$ which cannot be accounted for by \thi\ effects, but can be explained easily by H$_2$ formation. That CO-faint H$_2$ should be a significant component of the ISM is further supported by theoretical models \citep[e.g.,][]{2010ApJ...716.1191W}. Our results support these conclusions as well.

\section{Summary}
\label{s:summary}

To quantify the contribution of optically-thick gas to the ISM mass budget over large areas, we reproduce the F15 model for \hi\ properties (\thi\ and $T_s$) based on dust emission from \emph{Planck} and $21\rm\,cm$ emission from GALFA-\hi. We include additional masking for regions dominated by molecular gas based on $^{12}$CO $J=1-0$ maps from \emph{Planck}. Using this model, we compute the inferred correction to the \hi\ column density in the optically-thin limit (\Rdust). For comparison, we compute the correction to the optically-thin column density based on direct measurements of \thi\ (\Rhi) for a large sample of \thi\ observations (\totalLOS{} LOS), \galfaLOS{} of which are within in the unmasked GALFA-\hi\ FOV. Our results are as follows:

\begin{itemize}

\item Although \Rdust\ and \Rhi\ are consistent at low values, for all \galfaLOShigh{} LOS with \Rdust$>1.3$ we find significantly lower \Rhi\ ($1.0<$\Rhi$<1.3$). 
\item We rule out the possibility that our \thi\ LOS miss high optical-depth ``blobs" with small covering fraction, as these structures must have properties which are significantly incompatible with ISM conditions.
\item We conclude that the discrepancy between \Rhi\ and \Rdust\ rules out the F15 hypothesis that optically-thick \hi\ dominates the dark gas in the local ISM. Although we cannot distinguish here between intrinsic variations in dust grain emissivity and H$_2$ undetected by CO, both likely contribute significantly to the inferred dark gas mass budget.
\end{itemize}

\acknowledgements{We would like to thank the anonymous referee for thoughtful comments and suggestions which have improved this work.
We thank the participants of the $\Psi 2$ scientific program at Paris-Saclay University, ``The ISM Beyond 3D", for valuable discussions. 
This publication utilizes data from Galactic ALFA HI (GALFA-HI) survey data set obtained with the Arecibo L-band Feed Array (ALFA) on the Arecibo $305 \rm \, m$ telescope. The Arecibo Observatory is a facility of the National Science Foundation (NSF) operated by SRI International in alliance with the Universities Space Research Association and UMET under a cooperative agreement.  The GALFA-HI surveys are funded by the NSF through grants to Columbia University, the University of Wisconsin, and the University of California. This work makes use of data from the Karl G. Jansky Very Large Array, operated by the National Radio Astronomy Observatory (NRAO). NRAO is a facility of the NSF operated under cooperative agreement by Associated Universities, Inc. We acknowledge the use of the Legacy Archive for Microwave Background Data Analysis (LAMBDA), part of the High Energy Astrophysics Science Archive Center (HEASARC). HEASARC/LAMBDA is a service of the Astrophysics Science Division at the NASA Goddard Space Flight Center. This research has made use of NASA's Astrophysics Data System. This research made use of Astropy, a community-developed core Python package for Astronomy \citep{2013A&A...558A..33A}, NumPy \citep{van2011numpy}, and matplotlib, a Python library for publication quality graphics \citep{Hunter:2007}. }

\bibliography{ms}

\appendix{

To compare our results with those of F15, who used \hi\ data from the LAB survey \citep{2005A&A...440..775K}, and to ensure that our analysis is not biased by the GALFA-\hi\ FOV (i.e., $0<\alpha_{2000}<360^{\circ}$, $0<\delta_{2000}<35^{\circ}$), we repeat our analysis using $21\rm\,cm$ emission maps from the LAB survey, with all other maps at LAB survey resolution ($N_{\rm side}=128$, corresponding to $27.5^\prime$ pixels). An all-sky map of \Rdust\ (Equation~\ref{e:r_dust}) is shown in Figure~\ref{f:r_dust_map_lab}, with the positions of all \totalLOS{} \thi\ LOS overlaid as crosses, colored by \Rhi\ (Equation~\ref{e:r_hi}). We find excellent agreement with F15 (their Figure 12, lower panel), indicating that we are successfully reproducing their analysis method here.  We note that in comparison with F15, we have included additional masking at intermediate latitudes due to CO detected by \emph{Planck}, and we have not applied by-hand masking of the Magellanic System or intermediate-velocity gas, resulting in less masking at the highest Galactic latitudes. 

\begin{figure*}[t!]
\begin{center}
\includegraphics[width=\textwidth]{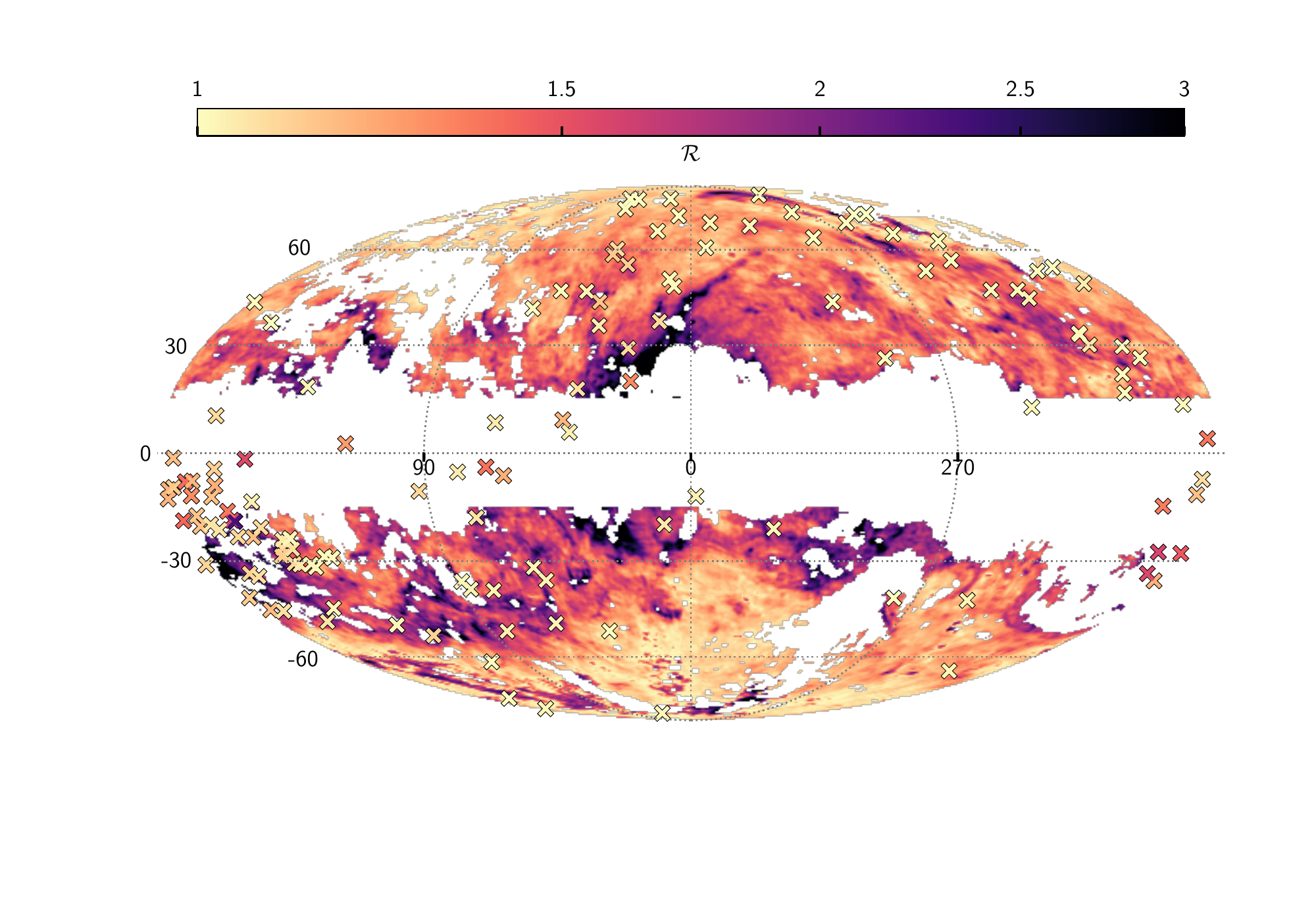}
\vspace{-70pt}
\caption{Map of \Rdust\ (Equation~\ref{e:r_dust}) for the full LAB sky, masked according to Section~\ref{s:mask}.  Pixels with undefined \Rdust\ (i.e., $\tau_{\rm dust} \leq 0.2$) are also masked. The positions of the \totalLOS{} \thi\ LOS are overlaid as crosses with colors corresponding to \Rhi\ (Equation~\ref{e:r_hi}).}
\end{center}
\label{f:r_dust_map_lab}
\end{figure*}
}

\end{document}